# 3D oxygen vacancy order and defect-property relations in multiferroic $(LuFeO_3)_9/(LuFe_2O_4)_1$ superlattices

K. A. Hunnestad[1], H. Das[2], C. Hatzoglou[1], M. Holtz[3,4], C. M. Brooks[4], A. T. J. van Helvoort[5], D. A. Muller[3,6], D. G. Schlom[4,6,7], J. A. Mundy[8], D. Meier[1*]

[1]Department of Materials Science and Engineering, NTNU Norwegian University of Science and Technology, 7491 Trondheim, Norway
[2]Institute of Innovative Research, WRHI, Tokyo Institute of Technology, 4259 Nagatsuta, Midori-ku Yokohama 226-8503, Japan
[3]School of Applied and Engineering Physics, Cornell University, Ithaca, New York 14853, USA
[4]Department of Materials Science and Engineering, Cornell University, Ithaca, New York 14853, USA
[5]Department of Physics, NTNU Norwegian University of Science and Technology, 7491 Trondheim, Norway
[6]Kavli Institute at Cornell for Nanoscience, Ithaca, New York 14853, USA
[7]Leibniz-Institut für Kristallzüchtung, Max-Born-Str. 2, 12489 Berlin, Germany
[8]Department of Physics, Harvard University, Cambridge, Massachusetts 02138, USA
*Corresponding author: dennis.meier@ntnu.no

Oxide heterostructures exhibit a vast variety of unique physical properties. Examples are unconventional superconductivity in layered nickelates[1] and topological polar order in $(PbTiO_3)_n/(SrTiO_3)_n$ superlattices[2,3]. Although it is clear that variations in oxygen content are crucial for the electronic correlation phenomena in oxides[4], it remains a major challenge to quantify their impact[5]. Here, we measure the chemical composition in multiferroic $(LuFeO_3)_9/(LuFe_2O_4)_1$ superlattices, revealing a one-to-one correlation between the distribution of oxygen vacancies and the electric and magnetic properties. Using atom probe tomography, we observe oxygen vacancies arranging in a layered three-dimensional structure with a local density on the order of $10^{14}$ cm$^{-2}$, congruent with the formula-unit-thick ferrimagnetic $LuFe_2O_4$ layers. The vacancy order is promoted by the locally reduced formation energy and plays a key role in stabilizing the ferroelectric domains and ferrimagnetism in the $LuFeO_3$ and $LuFe_2O_4$ layers, respectively. The results demonstrate the importance of oxygen vacancies for the room-temperature multiferroicity in this system and establish an approach for quantifying the oxygen defects with atomic-scale precision in 3D, giving new opportunities for deterministic defect-enabled property control in oxide heterostructures.

The concentration and distribution of oxygen in strongly correlated electron systems is essential for the material's response[5]. By introducing oxygen vacancies or interstitials, electronic and magnetic properties can be controlled, and even entirely new functional properties can be obtained[6]. For example, redox reactions can change the oxygen-stoichiometry and drive topotactic transitions[7], resistive switching[8], and ferroelectric self-poling[9]. In structured materials, the oxygen diffusion length is usually comparable to dimensions of the system[10] and local variations in oxygen content naturally arise due to varying defect formation energies[11]. The latter plays a crucial role for property-engineering in oxide heterostructures, where atomically precise interfaces in combination with defect engineering are used to tailor, e.g., polar order[12], magnetic exchange interactions[13], and the onset of superconductivity[14,15].

Quantifying emergent spatial variations in oxygen content at the atomic level, however, is extremely challenging[5,16]. Enabled by the remarkable progress in high-resolution transmission electron microscopy, it is possible to image individual oxygen defects in heterostructures[17] and, for sufficiently high defect densities, chemical fingerprints associated with their accumulation or depletion at interfaces/interlayers can be detected[18–21]. Despite their outstanding capabilities, these electron-microscopy based methods are not quantitative and inherently restricted to 2D projections along specific zone axes. This restriction prevents the full three-dimensional (3D) analysis of oxygen defects and limits the microscopic understanding of the interplay between oxygen defects and the material's physical properties. An experimental approach that, in principle, facilitates the required chemical accuracy and sensitivity to overcome this fundamental challenge is atom probe tomography (APT). The potential of APT is demonstrated by previous work on bulk oxide superconductors[4] and ferroelectrics[22], measuring stoichiometric variations at the nanoscale and lattice positions occupied by individual dopant atoms, respectively.

Here, we quantify the distribution of oxygen vacancies in $(LuFeO_3)_9/(LuFe_2O_4)_1$ superlattices and demonstrate its importance for the electric and magnetic orders that lead to room-temperature multiferroicity in this system. Using APT, we show that oxygen vacancies ($v_O$) have a propensity to accumulate in the $LuFe_2O_4$ monolayers, forming a layered 3D structure with an average density of about $(7.8 \pm 1.8) \cdot 10^{13}$ cm$^{-2}$. The oxygen vacancies facilitate the electrical screening that is essential for stabilizing the ferroelectric order and control the oxidation state of the iron (Fe), which is responsible for the emergent ferrimagnetism. The results clarify the defect–property relation and show that the multiferroic behavior in $(LuFeO_3)_9/(LuFe_2O_4)_1$ is intertwined with – and promoted by – the 3D oxygen vacancy order.

Figure 1a shows a high-angle annular dark-field scanning transmission electron microscopy (HAADF-STEM) image of the $(LuFeO_3)_9/(LuFe_2O_4)_1$ superlattice. The system exhibits spontaneous electric and magnetic order, facilitating magnetoelectric multiferroicity at room temperature[23]. The ferroelectricity relates to the displacement of the Lu atoms in the $LuFeO_3$ layers (up-up-down: +P; down-down-up: -P, see Fig. 1a), whereas the ferrimagnetism has been explained based on $Fe^{2+}$ → $Fe^{3+}$ charge-transfer excitations in the $LuFe_2O_4$ layers[24]. Interestingly, the multiferroic $(LuFeO_3)_9/(LuFe_2O_4)_1$ superlattice develops an unusual ferroelectric domain state with extended positively charged domain walls in the $LuFeO_3$ layers, where the polarization meets head-to-head (→←)[25]. The formation of charged head-to-head domain walls is surprising as they have high electrostatic costs, which raises the question how the material stabilizes them.

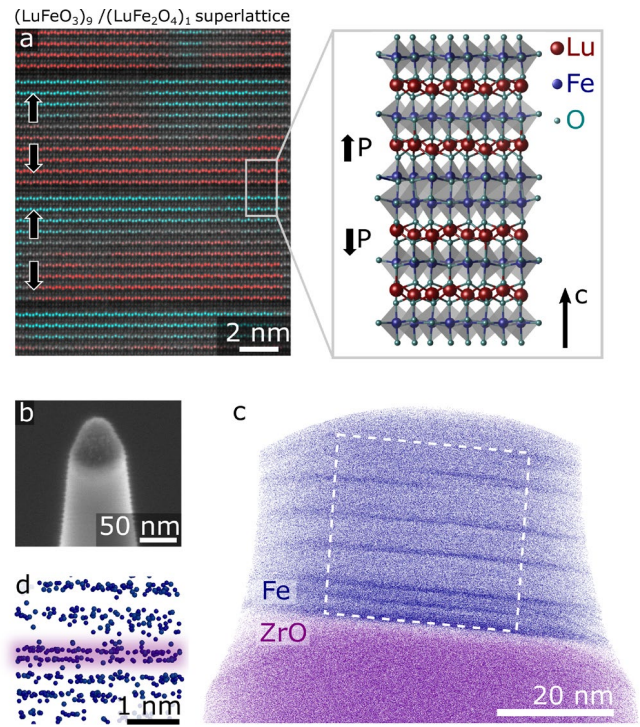

**Figure 1: 3D imaging of the $(LuFeO_3)_9/(LuFe_2O_4)_1$ superlattices structure.** **a**, HAADF-STEM image recorded along the [100] zone axis and schematic showing the atomic structure of the superlattice. Ferroelectric +P and -P domains are colored blue and red, respectively. **b**, SEM image of an APT needle. Three different layers are visible, corresponding to the Cr protection layer (dark grey), the $(LuFeO_3)_9/(LuFe_2O_4)_1$ superlattice (bright), and the substrate. **c**, 3D reconstruction of the APT data. Superlattice and substrate are represented by the Fe and ZrO ionic species, respectively. The dark lines in the superlattice correspond to double-Fe columns of the $LuFe_2O_4$ layers. **d**, Zoom-in to one of $LuFe_2O_4$ layers in **c**, resolving the double-Fe columns.

To understand the microscopic mechanism that leads to the distinct magnetic and electric order in $(LuFeO_3)_9/(LuFe_2O_4)_1$, we map the 3D chemical composition of the superlattice using APT. For the APT analysis, we deposit a protective capping layer (Pt, Cr or Ti) and prepare needle-shaped specimens using a focused ion beam (FIB, see Methods) as shown in Fig. 1b. The needle-like shape is a requirement in APT experiments and allows for producing the high electric fields required for field evaporation of surface atoms when a voltage > 2 kV is applied. The capping layer ensures that the (LuFeO$_3$)$_9$/(LuFe$_2$O$_4$)$_1$ superlattice is located below the tip of the needle, which enables us to analyze a larger volume and, hence, improve the chemical precision of the experiment. Figure 1c shows the 3D reconstruction of the measured volume, where Fe and ZrO ionic species are presented to visualize the superlattice and substrate, respectively (mass spectrum and bulk chemical composition are presented in Supplementary Fig. S1). The LuFe$_2$O$_4$ layers are visible as darker lines due to their higher concentration of Fe atoms compared to LuFeO$_3$. The 3D reconstruction shows that the spacing between the LuFe$_2$O$_4$ layers varies within the analyzed volume of the superlattice, ranging from approximately 2 nm to 6 nm. At the atomic scale, the LuFe$_2$O$_4$ layers exhibit the characteristic double-Fe layers (Fig. 1d), consistent with the HAADF-STEM data in Fig. 1a. Furthermore, enabled by the 3D APT imaging, we observe step-like discontinuities in individual LuFe$_2$O$_4$ layers in Fig. 1c. The observation of such growth-related imperfections leads us to the conclusion that the multiferroic response of the material is rather robust and resilient against such local disorder.

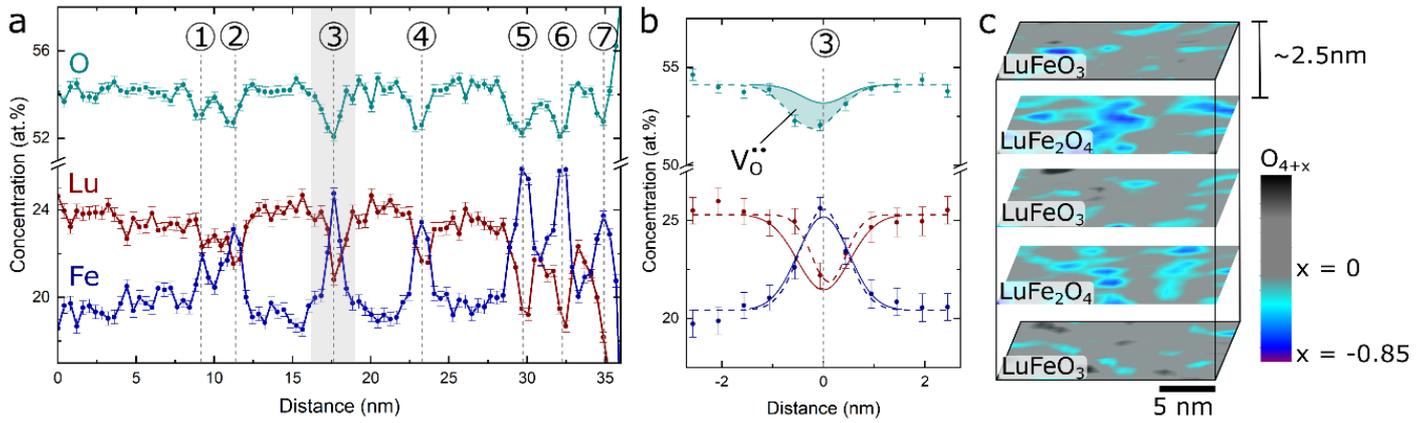

**Figure 2 | 3D oxygen vacancy order. a**, Profiles of the relative chemical composition, with the surface to the left in the plot. Anomalies are observed at all the LuFe$_2$O$_4$ layers, numbered ① to ⑦. **b**, Measured (data points) and theoretically expected (solid line) chemical concentration profile at LuFe$_2$O$_4$ layer ③. The shaded area highlights that the measured oxygen content is lower than in stoichiometric LuFe$_2$O$_4$, indicating an accumulation of oxygen vacancies. **c**, 3D visualization of the oxygen stoichiometry based on the APT data set. Oxygen vacancies arrange in a layered three-dimensional structure congruent with the formula-unit-thick ferrimagnetic LuFe$_2$O$_4$ layers. Within the LuFe$_2$O$_4$ layers, oxygen vacancies form puddle-like regions of reduced LuFe$_2$O$_{4-\delta}$ (blue).

Most importantly for this work, the APT measurement provides information about the local chemical composition of the superlattice. Figure 2a displays the concentration of the different atomic species evaluated for the region marked by the white dashed line in Fig. 1c. The line plots are derived by integrating the data in the direction perpendicular to the long axis of the needle-shaped sample, showing pronounced anomalies at the position of the LuFe$_2$O$_4$ layers (marked by dashed lines). In total, seven peaks are resolved labelled ① to ⑦; two peaks correspond to the discontinuous LuFe$_2$O$_4$ layer (represented by the double-peak ①/②) and five peaks to the continuous LuFe$_2$O$_4$ layers resolved in Fig. 1c (③ to ⑦). In all cases, we consistently find an enhancement in Fe concentration and a decrease in Lu and O concentration in the LuFe$_2$O$_4$ layers relative to LuFeO$_3$. A more detailed analysis of the chemical composition of one of the continuous LuFe$_2$O$_4$ layers (i.e., layer ③) is presented in Fig. 2b. Figure 2b compares measured and calculated concentration profiles for Lu, Fe, and O. The calculated concentration profile corresponds to a stoichiometric (LuFeO$_3$)$_9$/(LuFe$_2$O$_4$)$_1$ superlattice, assuming a realistic experimental resolution of about 0.6 nm, showing a good agreement with the experimental data for Lu and Fe. In contrast, the measured concentration of O is lower than expected, indicating an accumulation of oxygen vacancies, $v_O$. By integrating over the layer, we find a $v_O$ density of $(7.8 \pm 1.8) * 10^{13}$ cm$^{-2}$, which corresponds on average to an oxygen deficient state in LuFe$_2$O$_{4-\delta}$ with $\delta \approx 0.5$.

The same trend is observed for other LuFe$_2$O$_4$ layers with minor layer-to-layer variations in the $v_O$ density (see Supplementary Fig. S2), indicating that the oxygen vacancies form a layered 3D structure within the (LuFeO$_3$)$_9$/(LuFe$_2$O$_4$)$_1$ superlattice, congruent with the arrangement of the LuFe$_2$O$_4$ layers. It is important to note, however, that within the different layers the distribution of $v_O$ is inhomogeneous as shown by the 3D map in Fig. 2c. This map presents the local chemical composition and reflects the periodic variation in $v_O$ density in the LuFeO$_3$ and LuFe$_2$O$_4$ layers, consistent with the integrated data in Fig. 2a,b. Furthermore, it reveals a puddle-like distribution of the oxygen vacancies with puddle sizes in the order of a few nanometers and a maximum local $v_O$ density of up to $\approx 10^{14}$ cm$^{-2}$ (i.e., a reduction to LuFe$_2$O$_{3.25}$).

To better understand the propensity of the oxygen vacancies to accumulate at the LuFe$_2$O$_4$ layers, we calculate and compare the $v_O$ defect formation energies for LuFeO$_3$ and LuFe$_2$O$_4$ using density functional theory (DFT) calculations (Methods). Possible vacancy sites are located in the Lu- or Fe-layers ($v_O^{LuO_2}$ or $v_O^{FeO}$) as illustrated in Fig. 3a,b. The respective defect formation energies as function of temperature are plotted in Fig. 3c, showing that the formation energy for oxygen vacancies at $v_O^{FeO}$ sites is in general lower than at $v_O^{LuO_2}$ sites. In addition, the comparison of the data for LuFe$_2$O$_4$ and LuFeO$_3$ indicates that it is energetically favorable to accommodate oxygen vacancies in LuFe$_2$O$_4$, yielding an energy reduction of 0.5 eV per $v_O^{FeO}$ relative to oxygen vacancies in LuFeO$_3$. Thus, by accumulating oxygen vacancies in the LuFe$_2$O$_4$ layers, the (LuFeO$_3$)$_9$/(LuFe$_2$O$_4$)$_1$ superlattice can substantially reduce its energy, which promotes the formation of $v_O^{FeO}$-rich LuFe$_2$O$_4$ layers and, hence, a layered 3D v$_O$ order consistent with the APT results.

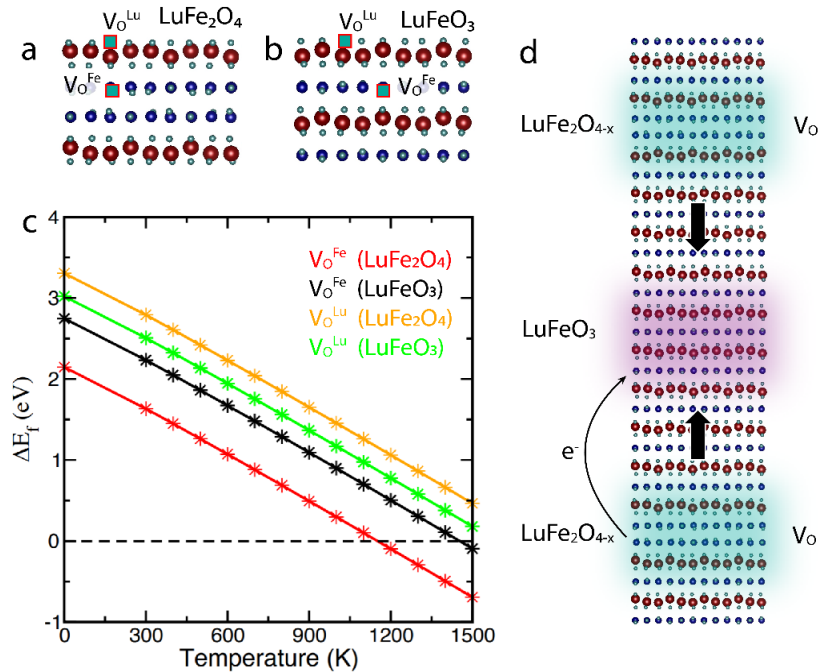

**Figure 3 | Defect formation energy for oxygen vacancies in LuFe$_2$O$_4$ and LuFeO$_3$. a,b**, Schematics illustrating possible defect sites for oxygen vacancies in LuFe$_2$O$_4$ and LuFeO$_3$, respectively. **c**, Comparison of oxygen vacancy formation energy as a function of temperature for an oxygen partial pressure of 10$^{-10}$ atm. The lowest energy is found for oxygen vacancies in the Fe layers of LuFe$_2$O$_4$ ($v_O^{FeO}$). **d,** Schematic of the superlattice structure, summarizing the APT and DFT results. Oxygen vacancies accumulate in the LuFe$_2$O$_4$ layers due to the locally reduced formation energy. The oxygen vacancies stabilize a ferroelectric tail-to-tail configuration at these layers and provide the electrons that are needed to screen the head-to-head domain walls that form in the LuFeO$_3$ layers.

The observed 3D oxygen vacancy order has a direct impact on electric and magnetic properties and provides insight into their microscopic origin. The accumulation of v$_O$ effectively leads to electron-doping of the LuFe$_2$O$_4$ layers. We find that the locally measured v$_O$ density (Fig. 2) is equivalent to a positive charge of $25 \pm 5$ μC/cm$^2$. The latter explains why the (LuFeO$_3$)$_9$/(LuFe$_2$O$_4$)$_1$ superlattice develops the unusual ferroelectric tail-to-tail configuration at the LuFe$_2$O$_4$ layers (seen in Fig. 1a and 3d). The latter carry a negative charge of about 12 μC/cm$^2$, which partially compensates the positive charge associated with the oxygen vacancies. As a consequence of the energetically favored tail-to-tail configuration at the LuFe$_2$O$_4$ layers, formation of positively charged head-to-head domain walls within the LuFeO$_3$ layers is enforced which, in turn, are screened by a redistribution of the mobile electrons generated by the v$_O$. The importance of this redistribution of electrons, i.e., the charge transfer of free electrons to the head-to-head domain walls goes beyond the ferroelectric properties. It also drives the change in the oxidation state of Fe in the LuFe$_2$O$_4$ layers (Fe$^{2+}$ → Fe$^{3+}$)[23–25], which is crucial for the ferrimagnetic order of the material.

In conclusion, both the electric and magnetic properties of (LuFeO$_3$)$_9$/(LuFe$_2$O$_4$)$_1$ are closely related to oxygen vacancy ordering. The results clarify the microscopic origin the unusual ferroelectric domain structure and the enhancement of the magnetic response, revealing the importance of extrinsic defect-driven mechanisms for the emergence of room-temperature multiferroicity. Quantitative 3D imaging of oxygen defects and chemical profiling at the atomic-scale is of interest beyond the defect-property relations discussed in this work and can provide insight into defect-driven effects and the general role that oxygen vacancies (or interstitial) play in emergent phenomena in oxide hetero-structures. The approach shown demonstrates the benefit of quantitative atomic-scale characterization of oxygen at interfaces in oxides, which is crucial for a better understanding of their complex chemistry and physics, as well as improved property engineering and their utilization in nanoelectronic and oxitronic technologies.

## Methods

**Sample preparation and characterization:** Thin films of (LuFeO$_3$)$_9$/(LuFe$_2$O$_4$)$_1$ were grown by reactive-oxide molecular-beam epitaxy in a Veeco GEN10 system on (111) (ZrO$_2$)$_{0.905}$(Y$_2$O$_3$)$_{0.095}$ (or 9.5 mol% yttria-stabilized zirconia) (YSZ) substrates, as described in Ref. [23]. A 300 nm Ti or Cr protective layer was deposited on top of the film with e-beam evaporation using a Pfeiffer Vacuum Classic 500, at a rate of 1 Å/s. The characteristic needle shaped specimens for APT were prepared with a Helios NanoLab DualBeam FIB as described by Ref. [26]. Cross-sectional TEM specimens were prepared using an FEI Strata 400 FIB with a final milling step of 2 keV to reduce surface damage.

**Transmission electron microscopy**: Selected specimens were inspected to ensure adequate sample quality with TEM using a JEOL JEM-2100F Field Emission Electron Microscope operating at 200kV. The high-resolution HAADF-STEM image in Fig. 1 was acquired on a 100-keV Nion UltraSTEM, a fifth-order aberration-corrected microscope. The lutetium distortions were quantified from HAADF-STEM images, as described in Ref. [23].

**Atom probe tomography**: APT measurements were recorded with a Cameca LEAP 5000XS instrument, operating in laser pulsed mode. Data was collected at cryogenic temperature ($T$ = 25 K) with an applied bias between 2 kV and 10 kV. Laser pulses with 30 pJ energy and 250 kHz frequency were used, and the detection rate was set to 0.5%, i.e., 2 ions detected every 1000 pulse. The raw APT data was reconstructed into 3D datasets with the Cameca IVAS 3.6.12 software, using the voltage profile to determine the radial evolution. The image compression factor and field reduction factor was adjusted to make the thin film flat relative to the substrate.

**First-principles calculations of oxygen vacancy formation.** To understand the tendency of formation of an oxygen vacancy (v$_O$) in the (LuFeO$_3$)$_9$/(LuFe$_2$O$_4$)$_1$ superlattice, we studied the formation energy ($\Delta E_f$) of an oxygen vacancy as a function of temperature ($T$) and oxygen partial pressure ($p$) by considering the bulk ferroelectric state of LuFeO$_3$ (space group $P6_3cm$) and the bulk ferroelectric Fe$^{2+}$/Fe$^{3+}$ charge-ordered state of LuFe$_2$O$_4$ (space group $Cm$). This is a reasonable consideration as in the (LuFeO$_3$)$_9$/(LuFe$_2$O$_4$)$_1$ superlattices the improper ferroelectric signature trimer distortions are induced in the LuFe$_2$O$_4$ layer by the LuFeO$_3$ layers[23]. The formation of oxygen vacancies was studied by extracting one oxygen atom from the supercell of the ferrite systems. We used the following equation to calculate $\Delta E_f$,[27,28]

$$\Delta E_f = E(v_O) - E_0 + \Delta x \mu_O$$

where $E(v_O)$ and $E_0$ represent the total energies of the supercell with and without an oxygen vacancy, respectively, and $\Delta x$ denotes the number of $v_O$ created in the supercell. As we considered a neutral oxygen vacancy, $\Delta E_f$ does not depend on the charge state of $v_O$ and the Fermi level of the system[28]. The chemical potential of oxygen atom, denoted as $\mu_O$, was calculated by the following equation[29],

$$\mu_O(p,T) = \mu_O(p_0,T_0) + \mu_O(p_1,T) + \frac{1}{2}k_B T \ln\left(\frac{p}{p_1}\right)$$

Here, $\mu_O(p_0,T_0)$ represents the oxygen chemical potential at zero pressure ($p_0$=0) and zero temperature ($T_0 = 0$). According to the first principles calculations, $\mu_O(p_0,T_0) = \frac{1}{2}E(O_2)$, where $E(O_2)$ denotes the total energy of an O$_2$ molecule. The second term, $\mu_O(p_1,T)$, which denotes the contribution of the temperature to the oxygen chemical potential at a particular pressure of $p_1 = 1$ atm, was obtained from the experimental data[30]. The third term, $\frac{1}{2}k_B T \ln\left(\frac{p}{p_1}\right)$, represents the contribution of pressure to the chemical potential of oxygen. Here, $k_B$ is the Boltzmann constant. In the present study, we considered two kinds of oxygen vacancies, located in the Lu- ($v_O^{LuO_2}$) or Fe ($v_O^{FeO}$) layers, as illustrated in Fig. 3a,b. The oxygen vacancy formation was calculated in supercells consisting of 12 formula units and 24 formula units of LuFe$_2$O$_4$ and LuFeO$_3$, respectively.

**Computational details.** We calculated the total energies by performing first-principles calculations by employing the density functional theory (DFT) method and the projector augmented plane-wave (PAW) basis method as implemented in the VASP (Vienna *Ab initio* Simulation Package)[31,32]. The Perdew-Burke-Ernzerhof (PBE) form of the generalized gradient approximation (GGA) was used to calculate the exchange correlation functional[33]. A kinetic energy cut-off value of 500 eV and appropriate k-point meshes were selected so that total ground state energies were converged to $10^{-6}$ eV and the Hellman-Feynman forces were converged to 0.001 eV Å$^{-1}$. For each structure, coordinates of all atoms and lattice vectors were fully relaxed. The GGA+U method as developed by Dudarev *et al.*[34] was employed to deal with electron correlation in the Fe 3*d* state. All calculations were performed by considering $U_{eff} = U - J_H = 4.5$ eV for the Fe 3*d* states, where $U$ and $J_H$ represent the spherically averaged matrix elements of on-site Coulomb interactions. Additionally, we cross-checked the value of $\Delta E_f$ by varying the value of $U_{eff}$ from 3.5 to 5.5 eV, as was used in the previous studies [23,35,36]. We considered Lu 4*f* states in the core. All the calculations of total energies were performed with ferromagnetic collinear arrangement of Fe spins and without spin-orbit coupling.

**Estimation of O vacancy density:** Due to a change in unit cell composition at the LuFe$_2$O$_4$ layer, the O vacancy density cannot directly be extracted from the profile in Fig. 2. Instead, a simulation based on the ideal superlattice structure without any defect was made (solid line in Fig. 2). Using a DFT-based structure of the superlattice, the ideal atomic distribution was simulated. The atoms were then shifted around randomly to simulate the spatial resolution of experiment, which was done with a gaussian distribution with 0.55 nm standard deviation. A chemical profile across the simulated structure was then performed to get an expectation of an ideal superlattice structure. The difference between the simulated profile to the real data (shaded area in Fig. 2) represents a measure for the $v_O$ concentration. This concentration was converted into a $v_O$ density by multiplying it with the oxygen density of the simulated data so that the limited APT detection efficiency is not affecting the final value. The 3D map of the oxygen depletion (Fig. 2c) is derived by displaying the chemical composition in the lateral dimension within five 20 x 20 x 1.5 nm$^3$ volumes. Chemical composition is converted into formula units (i.e., LuFe$_2$O$_x$) by measuring the local chemical composition, and compensating for the spatial resolution of the instrument as the oxygen depletion is spread out of regions larger than the actual LuFe$_2$O$_4$ layer.

## Data availability

Computer codes used for simulations and data evaluation are available from the sources cited. Other data in formats not presented within the paper are available from the corresponding author upon request.

## Acknowledgements

The Research Council of Norway is acknowledged for the support to the Norwegian Micro- and Nano-Fabrication Facility, NorFab, project number 295864, the Norwegian Laboratory for Mineral and Materials Characterisation, MiMaC, project number 269842/F50, and the Norwegian Center for Transmission Electron Microscopy, NORTEM (197405/F50). K.A.H. and D.M. thank the Department of Materials Science and Engineering at NTNU for direct financial support. D.M. acknowledges funding from the European Research Council (ERC) under the European Union's Horizon 2020 research and innovation program (Grant Agreement No. 863691) and further thanks NTNU for support through the Onsager Fellowship Program and NTNU Stjerneprogrammet. Research at the Tokyo Institute of Technology is supported by the Grants-in-Aid for Scientific Research Grants No. JP19K05246 and No. JP19H05625 from the Japan Society for the Promotion of Science (JSPS). H.D. also acknowledges computational support from TSUBAME supercomputing facility. Film growth and electron microscopy characterization is supported by the US Department of Energy, Office of Basic Energy Sciences, Division of Materials Sciences and Engineering, under Award No. DE-SC0002334. The Electron Microscopy Facilities at the Cornell Center for Materials are supported through the NSF MRSEC program (DMR-1719875).

## Author contributions

K.A.H conducted the APT experiments, including sample preparation and data analysis, supported by C.H., under the supervision of A.T.J.v.H and D.M. Samples by C.M.B., D.G.S., and J.A.M. and atomic-STEM by M.E.H. and D.A.M. DFT calculations were performed by H.D. D.M. initiated and coordinated the project. K.A.H. and D.M. wrote the manuscript. All authors discussed the results and contributed to the final version of the manuscript.

Supplementary information

# 3D oxygen vacancy order and defect-property relations in multiferroic $(LuFeO_3)_9/(LuFe_2O_4)_1$ superlattices

K. A. Hunnestad[1], H. Das[2], C. Hatzoglou[1], M. Holtz[3,4], C. M. Brooks[4], A. T. J. van Helvoort[5], D. A. Muller[3,6], D. G. Schlom[4,6,7], J. A. Mundy[8], D. Meier[1]

[1]Department of Materials Science and Engineering, NTNU Norwegian University of Science and Technology, 7491 Trondheim, Norway
[2]Institute of Innovative Research, WRHI, Tokyo Institute of Technology, 4259 Nagatsuta, Midori-ku Yokohama 226-8503, Japan
[3]School of Applied and Engineering Physics, Cornell University, Ithaca, New York 14853, USA
[4]Department of Materials Science and Engineering, Cornell University, Ithaca, New York 14853, USA
[5]Department of Physics, NTNU Norwegian University of Science and Technology, 7491 Trondheim Norway
[6]Kavli Institute at Cornell for Nanoscience, Ithaca, New York 14853, USA
[7]Leibniz-Institut für Kristallzüchtung, Max-Born-Str. 2, 12489 Berlin, Germany
[8]Department of Physics, Harvard University, Cambridge, Massachusetts 02138, USA

## Supplementary Figures

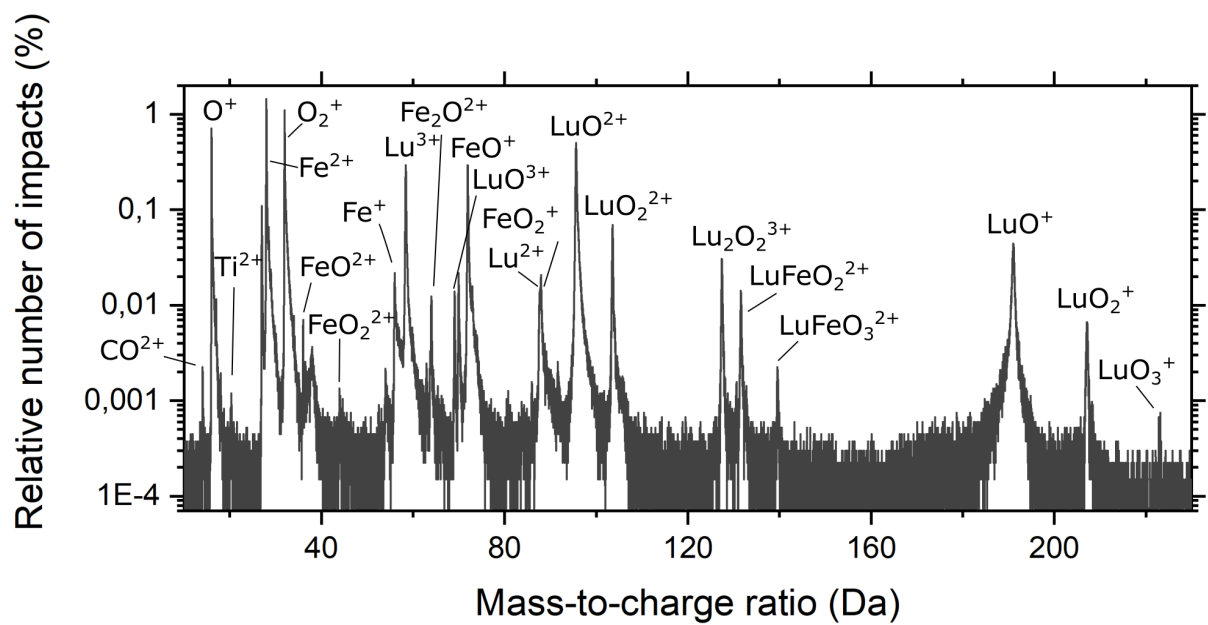

**Figure S1 | Mass spectrum of the (LuFeO$_3$)$_9$/(LuFe$_2$O$_4$)$_1$ superlattice.** Figure shows the mass spectrum of the superlattice shown in Fig. 1 in the main text. The spectrum is mainly coming from the thin film, but minor peaks from the substrate and capping layer are present, such as Ti$^{2+}$, which originates from the Ti protection layer. Species coming from the substrate are not labelled. In cases where peaks from the substrate and thin film overlap, the peak is assigned to the species belonging to the thin film. By integrating over the peaks, the average chemical composition from the thin film is found to be 53.74 ± 0.04 at.% O, 23.78 ± 0.05 at.% Lu and 22.56 ± 0.05 at.% Fe.

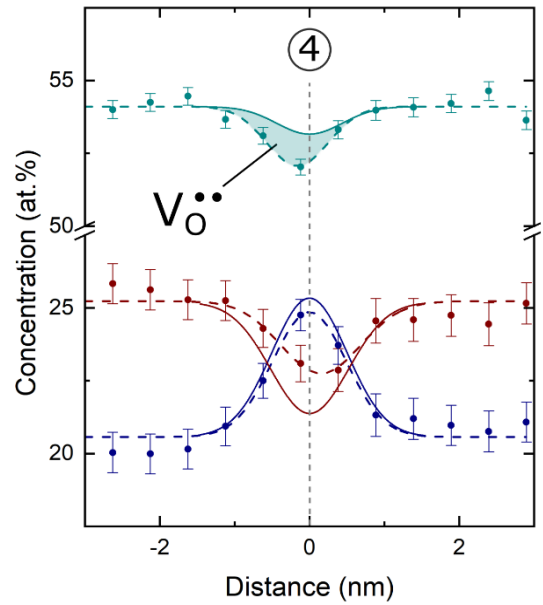

**Figure S2| APT analysis of a second double-Fe layer in the $(LuFeO_3)_9/(LuFe_2O_4)_1$ superlattices.** Peak ④ is shown (dashed line is a gaussian fit) and compared to a simulation (solid line) of the expected concentration profile from the superlattice. The deviation in O concentration (highlighted by the shaded area) indicates an oxygen vacancy density ($v_O$) of about $(5.6 \pm 1.7) \times 10^{13}/cm^2$.